\documentclass[a4paper,12pt]{article}
\usepackage[pctex32]{graphics}

\begin{document}
\topmargin 0pt \oddsidemargin 0mm

\renewcommand{\thefootnote}{\fnsymbol{footnote}}
\begin{titlepage}

\vspace{5mm}
\begin{center}
{\Large \bf Entropic force and its cosmological implications}
\vspace{12mm}

{\large  Yun Soo Myung\footnote{e-mail
 address: ysmyung@inje.ac.kr}}
 \\
\vspace{10mm} {\em  Institute of Basic Sciences and School of
Computer Aided Science, Inje University Gimhae 621-749, Korea}
\end{center}

\vspace{5mm} \centerline{{\bf{Abstract}}}
 \vspace{5mm}
We investigate a possibility of realizing the entropic force into
the cosmology.  A main issue is how the holographic screen is
implemented  in  the Newtonian cosmology. Contrary  to  the
relativistic realization of Friedmann equations, we do not clarify
the connection between   Newtonian cosmology and  entropic force
because there is  no way of implementing the holographic screen in
the Newtonian cosmology.

\end{titlepage}

\newpage
\renewcommand{\thefootnote}{\arabic{footnote}}
\setcounter{footnote}{0} \setcounter{page}{2}

%%===================section 1 ====================
\section{Introduction}
Since the discovery of the laws of black hole
thermodynamics~\cite{BCH}, Bekenstein~\cite{Bek} and
Hawking~\cite{Hawk} have suggested a deep connection between gravity
and thermodynamics, realizing black hole entropy and Hawking
radiation.  Later on, Jacobson~\cite{Jac} has demonstrated that
Einstein equations (describing relativistic gravitation) could be
derived by combining general thermodynamic pictures with the
equivalence principle. Padmanabhan~\cite{Pad3} has observed that the
equipartition law for  horizon degrees of freedom combined with the
Smarr formula leads to the Newton's law of gravity.

Recently, Verlinde has proposed the Newtonian force law as an
entropic force (non-relativistic version)  by using  the holographic
principle and  the equipartition rule~\cite{Ver}.
 If it
is proven correct, gravity is not a fundamental interaction, but an
emergent phenomenon which arises from the statistical behavior of
microscopic degrees of freedom encoded on a holographic screen. In
other words, the force of gravity is not something ingrained in
matter itself, but it is an extra physical effect, emerging from the
interplay of mass, time, space, and information.

After his work,  taking the apparent horizon as a holographic screen
(HS) to derive  the Friedmann equations~\cite{SG}, derivation of the
Friedmann equations using the equipartition rule and unproved Unruh
temperature~\cite{Pad1,CCO}, the modified equipartition rule to
discuss the large scale universe~\cite{Gao}, and the correction to
the entropic force by the corrected-entropy~\cite{shek} were
investigated for cosmological purpose.  The connection to the loop
quantum gravity~\cite{Smo}, the accelerating
surfaces~\cite{makela1}, holographic actions for black hole
entropy~\cite{CM}, and application to holographic dark
energy~\cite{LWh} were considered from the view of the entropic
force.   Furthermore,  cosmological implications were reported in
\cite{ZGZ,Wang,WLW,LW,LKL,EFS,Wei}, an extension to the Coulomb
force~\cite{WangT}, and the symmetry aspect of the entropy
force~\cite{Zhao} were investigated. The entropic force was
discussed in the presence of  black
hole~\cite{Myungef,LWW,TW,CCO2,PW}. The Schwarzschild spacetime was
introduced to  define the holographic screen properly~\cite{MKent}
and the entropic force did not always imply the Newtonian force law
when imposing the non-gravitational collapse
condition~\cite{Myungentn}.

However, one of urgent issues to resolve is to answer to the
question of how one can construct a spherical holographic screen of
radius $R$ which encloses a source mass $M$ located at the origin to
understand the entropic force.  This is a critical and important
issue because the holographic screen (an exotic description of
spacetime) originates from relativistic approaches to black hole
~\cite{Hoo,Suss} and cosmology~\cite{Bou}.  Verlinde has introduced
this screen  by analogy with an absorbing process of a particle
around the event horizon of black hole. Considering a smaller test
mass $m$ located at $\Delta x$ away from the screen and getting the
change of entropy on the  screen,  its behavior should resemble that
of a particle approaching a stretched horizon of a black hole, as
described by Bekenstein~\cite{Bek}.

Before proceeding, we would like to mention what is the difference
between Newtonian gravity and general relativity~\cite{NHw}. First
Newtonian gravity is an action-at-a-distance, that is, the
gravitational influence propagates instantaneously ($c\to \infty$),
implying the violation of causality. Second, Newtonian gravity is
ignorant of the presence of horizon where the relativistic effects
are supposed to dominates. For example, the horizons are  considered
as  either the event horizon of black hole or the apparent horizons
in the Friedmann-Robertson-Walker
 (FRW) universe.
Comparing Newtonian gravity and general relativity in cosmology is
different than in the case of isolated, asymptotically flat
systems~\cite{Raa}. For isolated systems, both Newtonian gravity and
general relativity are well-defined.  In contrast, while
relativistic cosmology is well-defined, there is no unique way to
accommodate Newtonian theory of cosmology because the Newtonian
equations are only defined up to boundary terms which have to be
specified at all times.

Hence it is not easy to implement the entropic force into the
cosmological setting.  In the literatures~\cite{SG,CCO,shek},  the
authors did not mention explicitly how the entropic force
(\ref{eq3}) works for the cosmological purpose.   It seems that the
entropic force is not realized in the Newtonian cosmology  unless
the holographic screen is clearly defined.

In this work, we investigate intensively how the Newtonian cosmology
is realized from the Poisson equation and Euler equation, the energy
consideration, and spherical cavity together with the cosmological
principle.  If the boundary surface enclosing the cavity filled with
the dust matter (or a source mass $M$) were replaced by the
holographic screen to which the equipartition rule and the
holographic principle are applied,  the entropic force  would derive
the evolution of the dust matter-dominated universe.

\section{Entropic force}

When a test particle with mass $m$ is close to a surface ${\cal S}$
(holographic screen) with distance $\Delta x$ (compared to the
Compton wave length $\lambda_m=\frac{\hbar}{mc}$), the change of
entropy on the holographic screen takes the form~\cite{Ver}
\begin{equation} \label{eq1} \Delta S=2\pi k_B\frac{\Delta
x}{\lambda_m} \to  2\pi m \Delta x
\end{equation} in the natural units of $\hbar=c=k_B=1$ and
$G=l^2_{pl}$. Considering that the entropy of a system depends on
the distance $\Delta x$, an entropic force $F_{ent}$ could be arisen
from the analogy of the biophysics  \begin{equation} \label{eq2}
F_{ent} \Delta x=T \Delta S
\end{equation} which may be  considered as an indication that the first
law of thermodynamics is realized on the holographic screen.
Plugging (\ref{eq1}) into (\ref{eq2}) leads to an important
connection between entropic force and temperature  on the
holographic screen
\begin{equation} \label{eq3} F_{ent}=2\pi m T. \end{equation} It implies that if one
knows the temperature $T$ on the holographic screen, the entropic
force is determined by (\ref{eq3}). Therefore, a key step is to
determine the temperature on the holographic screen.  Let us assume
that the energy $E$ is distributed on a spherical shape of
holographic screen with radius $R$ and  the mass $M$ is located at
the origin of coordinate as the source mass. Then, we may introduce
the equipartition rule to define the temperature
$T$~\cite{Pad2,Pad3}, the equality of energy and mass, and the
holographic principle to give the number of states $N$,
respectively, as
\begin{eqnarray}
\label{eq4e} E&=&\frac{1}{2 }N T,\\
\label{eq4m}E&=&M,\\
\label{eq4n}N&=&\frac{A}{G}\end{eqnarray} with the area of a
holographic screen $A=4\pi R^2$.  These are combined to determine
the temperature on the holographic screen \begin{equation}
\label{eq5} T=\frac{GM}{2\pi R^2}. \end{equation} Substituting
(\ref{eq5}) into (\ref{eq3}), the entropic force is realized as the
Newtonian force law \begin{equation} \label{eq6} F_{ent}=\frac{G m
M}{R^2}. \end{equation}

On the other hand, considering that Unruh has proposed the
connection between acceleration and temperature
\begin{equation}
T_U=\frac{a}{2\pi}\to T, \end{equation} Eq.(\ref{eq3}) leads to the
second law of Newton
\begin{equation}
F_{ent}=ma.
\end{equation}
We remind  the reader that $T_U$ is the bulk temperature, while $T$
is the boundary surface temperature.

\section{Newtonian cosmology from the Poisson equation}
We start with the Poisson  equation for the Newtonian potential
$\phi$~\cite{Rain}
\begin{equation} \phi_{,ii} = 4\pi G\rho \label{Poisson} \end{equation}
with $i=1,2,3$.
  The continuity
and Euler equations of fluid dynamics are given by
\begin{eqnarray}
\dot{\rho} + \rho v_{i,i} = 0, \label{Continuity} \\
\dot{v_i} + \phi_{,i} + \frac{1}{\rho}p_{,i} = 0, \label{Euler}
\end{eqnarray}
where $v_i$ is the velocity field and $p$ is the pressure.
Homogeneity implies that the density and pressure are merely
functions of time: $\rho(t)$ and $p(t)$. Also  the velocity field is
the same relative to all observers, which implies $v_i = V_{ij}(t)
x_j$. Substituting the Poisson equation into the Euler equation
leads to the fact that the Newtonian potential must take  the form
\begin{equation}
\phi = a_{ij}(t)x_ix_j + a(t). \end{equation} Hence, the Newtonian
approximation of a homogeneous cosmology is determined by
\begin{eqnarray}
 a_{ii} = 4 \pi G \rho,~\dot{\rho} + \rho V_{ii} = 0,~\dot{V_{ij}}
+ V_{ik}V_{kj} = a_{ij}. \label{nc}
\end{eqnarray}
Before we proceed, we mention that the relativistic
(Friedmann-Robertson-Walker: FRW) cosmology is based on the isotropy
and homogeneity. Thus, we will only consider the isotropic case
where (\ref{nc}) becomes shear-free and rotation-free. However, in
general, the Newtonian cosmology is anisotropic where there exist
shear and rotation. To this end, we consider the following SO(3)
decomposition
\begin{eqnarray}
V_{ij} = \frac{1}{3} \theta \delta_{ij} + \sigma_{ij} + w_{ij},
\label{decomposition}
\end{eqnarray}
where
\begin{eqnarray}
\theta = V_{ii},~ \sigma_{ij} =\frac{1}{2} (V_{ij} + V_{ji}) -
\frac{1}{3} \theta \delta_{ij},~ w_{ij} =\frac{1}{2}(V_{ij} -
V_{ji}). \label{tta}
\end{eqnarray}
Here, the trace part $\theta$ denotes the expansion, the trace-free
symmetric tensor  $\sigma_{ij}$ represents the shear,  and the
anti-symmetric tensor $w_{ij}$ describes the rotation. Plugging this
decomposition into the Euler equation, and setting $\sigma_{ij} = 0$
and $w_{ij} = 0$ for a shear-free and rotation-free fluid, we obtain
the equation for $\theta$ and the diagonalization of $a_{ij}$,
respectively,
\begin{eqnarray}
\dot{\theta} =  - \frac{1}{3}\theta^2 - 4\pi G\rho,~ a_{ij} =
\frac{1}{3}a_{kk}\delta_{ij}
\end{eqnarray}
with continuity equation
\begin{eqnarray}
\dot{\rho} + \rho \theta = 0. \label{cont}
\end{eqnarray}
Rewriting the expansion parameter $\theta =
3\frac{\dot{\tilde{R}}}{\tilde{R}}$ in terms of a Newtonian scale
factor $\tilde{R}$, the solution  to the continuity equation
(\ref{cont}) is
\begin{eqnarray}
\rho = \frac{C'}{ \tilde{R}^{3}}, \label{rho}
\end{eqnarray}
with $C'$ a constant.

Using (\ref{rho}) in the Euler equation, we see that the isotropic
and homogeneous Newtonian cosmology is described  by
\begin{eqnarray}
a_{ii} & = & 4\pi G\rho, \label{nc1} \\
\frac{\ddot{\tilde{R}}}{\tilde{R}} & = & - \frac{4}{3} \pi G \rho, \label{nc2} \\
\rho & = & \frac{C'}{ \tilde{R}^{3}}. \label{nc3}
\end{eqnarray}
At this stage, we mention the  FRW case  whose  equations are given
by the Friedmann equation without the tilde ( $\tilde{}$ ) notation
\begin{eqnarray}
\left(\frac{\dot{R}}{R}\right)^2  =  \frac{8}{3}\pi G
\rho-\frac{k}{R^2}, \label{Friedmann}
\end{eqnarray}
and  the Raychaudhuri equation
\begin{eqnarray}
\frac{\ddot{R}}{R} = - \frac{4\pi G}{3} \Bigg(\rho + 3p\Bigg),
\label{Ray}
\end{eqnarray}
from which one may obtain the Bianchi identity
\begin{eqnarray} \label{bian}
\dot{\rho} + 3\Bigg(\rho + p\Bigg)\frac{\dot{R}}{R} = 0.
\end{eqnarray}
We note  the difference that $R(t)$ is the relativistic scale factor
in (\ref{Friedmann})-(\ref{bian}), while  $\tilde{R}(t)$ is  the
Newtonian scale factor in the Newtonian cosmology
(\ref{nc1})-(\ref{nc3}).

 We are
now in a position to compare these two theories. The general
relativistic theory is well-posed. Equations (\ref{Friedmann}) and
(\ref{Ray}) are consistent with the Bianchi identity. In the
Newtonian theory there is only one equation (\ref{nc1}), and there
is no completeness because (\ref{nc1}) does not give (\ref{nc2}) and
(\ref{nc3}), and nor is the theory well-posed. Also, notice that in
the general relativistic theory, pressure occurs in the dynamics of
the theory, whereas in the Newtonian theory, pressure does not occur
anywhere in the dynamics and is only defined through an equation of
state.  Equation (\ref{nc2}) has the same form as the Raychaudhuri
equation (at least when $p = 0$).

Let us answer to the question of how do $R(t)$ and $\tilde{R}(t)$
differ. Considering
\begin{eqnarray}
a_{ii} & = & - 3 \frac{\ddot{\tilde{R}}}{\tilde{R}},~ a(t) =
\dot{A},
\end{eqnarray} the Poisson equation
(\ref{nc1}) yields
\begin{eqnarray}
\frac{\ddot{\tilde{R}}}{\tilde{R}} = - \frac{4}{3}\pi G
\rho.\label{nop}
\end{eqnarray}
This is again the Raychaudhuri equation of general relativity for
the dust matter $p=0$. Hence, the general relativistic scale factor
$R(t)$ is equivalent to the Newtonian scale factor $\tilde{R}(t)$.
However, it is argued that Newtonian cosmology is applicable only
when confined to a neighborhood of the observer, corresponding to
distances which are small compared to the Hubble distance $d_H=1/H$.
Therefore, {\it it is problematic to define the apparent horizon as
the holographic screen in the Newtonian cosmology}. For a flat
spacetime, the apparent horizon occurs at $r_A=1/H$.

Furthermore, the Newtonian theory suffers in that varying the
equation of state $p=\omega \rho$ will have no effect on the outcome
of the solutions for $\rho (t)$ and $\tilde{R}(t)$. This is because
the pressure does not appear in the dynamical equations.  Thus,  we
may at most reproduce the results of the matter-dominated case of
general relativity. Finally, it is worth to note  that the Friedmann
equation (\ref{Friedmann}) was missed in the Newtonian cosmology.
This equation could be realized from another approach of Newtonian
mechanics together with its energy consideration.

Consequently, the Euler equation (\ref{Euler}) leads to the
Raychaudhuri equation (\ref{nc2}) for the description of Newtonian
cosmology when combined with the cosmological principle.

\section{Newtonian cosmology from energy} In this
section, we will show how the Friedmann equation (\ref{Friedmann})
comes out from the Newtonian mechanics.   We propose that a system
of the universe consists of a number $N$ of galaxies with their mass
$m_p$ and position ${\bf r}_p(t)=r_p(t)\hat {\bf r} $ as measured
from a fixed origin O~\cite{Myungenergy}. Then  the kinetic energy
$T$ of the system  is given by
\begin{equation}
\label{2eq1} T= \frac{1}{2} \sum^{N}_{p=1} m_p \dot r^2_p.
\end{equation}
The gravitational potential energy $V$ is
\begin{equation}
\label{2eq2} V_g= -G \sum^{N}_{p<q} \frac{m_p m_q}{|{\bf r}_p-{\bf
r}_q|}.
\end{equation}
Then, the total energy $E$ of this system is given by
\begin{equation}
\label{2eq4} E=\frac{1}{2} \sum^{N}_{p=1} m_p \dot r^2_p -G
\sum^{N}_{p<q} \frac{m_p m_q}{|{\bf r}_p-{\bf r}_q|}.
\end{equation}
Suppose that the distribution and motion of the system is known at
some fixed epoch $t=t_0$ as an initial condition. Invoking the
cosmological principle of homogeneity and isotropy, the radial
motion at any time $t$ is then given by $r_p(t)=S(t)r_p(t_0)$ where
$S(t)$ is a universal function of time which is the same for all
galaxies and is related to the Newtonian scale factor. Substituting
this into Eq.(\ref{2eq4}) leads to an energy relation
\begin{equation}
\label{2eq5} E=A \dot S(t)^2 -G\frac{B}{S(t)},
\end{equation}
where  coefficients $A$ and $B$ are positive constants given by
\begin{equation}
\label{2eq6} A= \frac{1}{2} \sum^{N}_{p=1} m_p [r_p(t_0)]^2,~~ B=
\sum^{N}_{p<q} \frac{m_p m_q}{|{\bf r}_p(t_0)-{\bf r}_q(t_0)|}.
\end{equation}
Here $B$ contains the gravitational configuration of the system at
the initial time $t=t_0$. Eq.(\ref{2eq5}) is one form of the
cosmological differential equation for a scale factor $S(t)$. If the
universe is expanding, $A$-term decreases since the total energy
remains constant as $B$-term decreases. Therefore, the expansion
must slow down. Introducing the Newtonian scale factor with $S(t)=
\mu \tilde{R}(t)$, Eq.(\ref{2eq5}) takes the form
\begin{equation}
\label{2eq7}
\frac{\dot{\tilde{R}}^2}{\tilde{R}^2}=G\frac{C_1}{\tilde{R}^3}
-\frac{k}{\tilde{R}^2},
\end{equation}
where the constants $C_1$ and $k$ are defined by $C_1=\frac{B}{
\mu^3A}$ and $k=-\frac{E}{\mu^2A}$. When $E=0$, $\mu$ is arbitrary.
However, if $E \not=0$, one can choose $\mu^2=|E|/A$ so that
$k=1,0,-1$. That is, the sign of $E$ determines the evolution of the
universe: for $E<0$, it will contract, while for $E>0$ it will
expand.  This equation is the same form of the Friedmann equation
(\ref{Friedmann}) of a relativistic cosmology. We mention that there
exist ambiguities in determining the cosmological parameters $C_1$
and $k$. However, we note that the term in the left-hand side of
Eq.(\ref{2eq7}) originates from the kinetic energy,  the first term
(last term) in the right-hand side come from the potential energy
(total energy). In order to derive the Raychaudhuri equation, one
may use the conservation of energy $(\dot{E}=0)$ to find
\begin{equation} \label{newR}
\frac{\ddot{\tilde{R}}}{\tilde{R}}=-\frac{GC_1}{2}\frac{1}{\tilde{R}^3}.
\end{equation}
If $C_1=\frac{8\pi}{3}C'$, the above equation leads to the
Raychaudhuri equation (\ref{Ray}) with $p=0$. One may attempt to
interpret (\ref{newR}) as the Newtonian force equation for a test
mass with $m=1$ on a $S^2$ of a  proper radius of
$\hat{r}=r\tilde{R}(t)$. Here, $\tilde{R}$ is the Newtonian scale
factor to describe the evolution of the matter-dominated universe.
However,  it is not clear that (\ref{newR}) is interpreted as the
Newtonian force law.

Finally, we have derived the Friedmann equation (\ref{2eq7}) from
the energy condition together with cosmological principle.

\section{Newtonian force law on an expanding cavity}

Let us introduce an expanding cavity of radius $\hat{r}$ centered at
O whose volume is $V=4\pi \tilde{r}^3/3$. According to the Gauss
theorem in Newtonian mechanics (Birkoff's theorem in general
relativity), the net gravitational effects of a uniform external
medium on a spherical cavity is zero. In other words, the force
acting on a test mass located at the boundary of $\partial V=S^2$ is
the gravitational attraction from the matter internal to $\hat{r}$
only, which may be considered  as a point mass $M$ at O. This is
close to the situation to define the entropic force. {\it However,
we never choose  this boundary as the holographic screen.} Here, we
follow the classical approach to finding the Newtonian force law.
The mechanical energy of a test particle at the boundary is given by
the sum of kinetic and gravitational potential energy as
\begin{equation} \label{neq1}
U=\frac{1}{2} m \dot{\tilde{r}}^2-\frac{GmM}{\tilde{r}^2}
=\frac{1}{2} m \dot{\tilde{r}}^2-\frac{4\pi}{3} G m \rho
\tilde{r}^2,
\end{equation}
where the mass inside the cavity is
\begin{equation}
M=\rho V.
\end{equation}
Rewriting  the physical distance $\tilde{r}$  in terms of the
comoving distance $r$ and the Newtonian scale factor $\tilde{R}$ as
\begin{equation}
\tilde{r}=r \tilde{R}(t),
\end{equation}
the energy conservation law leads to
\begin{equation} \label{energyc}
U=\frac{1}{2} m \dot{\tilde{R}}^2 r^2-\frac{4\pi}{3} G m \rho
\tilde{R}^2 r^2.
\end{equation}
This can be rearranged into the Friedmann equation as
\begin{equation}
\frac{\dot{\tilde{R}}^2}{\tilde{R}^2}=\frac{8\pi G}{3}
\rho-\frac{\tilde{k}}{\tilde{R}^2}, \end{equation} where
\begin{equation}
\tilde{k}=-\frac{2U}{m r^2}
\end{equation}
which is the same form as $k$ in (\ref{2eq7}) when the
correspondence is assumed to be
\begin{equation}
E \longleftrightarrow U,~\mu \longleftrightarrow r,~ A
\longleftrightarrow \frac{m}{2}.
\end{equation}
Hence, we note that $\tilde{k}$ is constant, implying that $U
\propto r^2$. This means that while $U$ is constant for a given
particle, it changes if one looks at different comoving separations
$r$.

 Differentiating  the energy conservation (\ref{neq1})
with respect to time $t$ ($\dot{U}=0$), we have the Newtonian force
law on $m$ located at the boundary surface
\begin{equation} \label{NFL}
m \ddot{\tilde{R}}=-\frac{GmM}{\tilde{R}^2},
\end{equation}
which states clearly  that the only force on $m$ is due to the mass
$M$ inside the cavity. In deriving (\ref{NFL}), we used the constant
of mass $M$ in the cavity \begin{equation} \label{consM} \dot{M}=0.
\end{equation}
We note that dividing  the Newtonian force law (\ref{NFL}) by $m$
leads to the Raychaudhuri equation (\ref{newR}) which is an
accelerating equation for the Newtonian cosmology. This means that
in the Newtonian cosmology, the Newtonian force law determines the
acceleration of the cosmological evolution. One may consider that
(\ref{NFL}) is the Newtonian force law for a mass $m$ which is
circulating around the source mass $M$ because the left land side
seems to be a centripetal force. However, for cosmological purpose,
we consider the motion in the radial direction but not the motion in
the tangential direction.

At this stage, we introduce the first law  which takes the form
\begin{equation}
dE+pdV=TdS. \end{equation} We  apply the first law of thermodynamics
to an expanding cavity of unit comoving radius with $r=1$ filled by
$M$. Using $E=M$, the first law leads to \begin{equation} dM=TdS
\end{equation}
which for an isoentropic  expansion of $dS=0$, it leads to
(\ref{consM}) and finally
\begin{equation}
dM=0 \to \dot{M}=0 \to \dot{\rho}+3\frac{\dot{\tilde{R}}}{\tilde{R}}
\rho=0,
\end{equation} where the last expression is simply the continuity equation (\ref{cont}) for the
dust matter.

Consequently, we have found the Newtonian force law on the test
particle $m$ located at the boundary of an expanding  cavity from
the energy consideration together with the constant mass (continuity
equation).  We have considered regions smaller than the Hubble
horizon ($\tilde{r} \ll 1/H$) and  the expansion velocity are small
($v \ll c$) and thus, nonrelativistic dynamics are used.   In this
case, the first law of thermodynamics simply implies the continuity
equation for the dust matter.

\section{Entropic force and cosmological implications}
In order to see what happens when the entropic force was introduced
to describe the cosmology, let us consider a few  of relevant works.
For this purpose,  a radical  change should be made  such a way that
 the boundary surface enclosing the spherical cavity is replaced by the
holographic screen:
\begin{equation} \label{hsp}
{\rm boundary~surface}~(\partial {\cal V}) ~~ \longrightarrow~~ {\rm
holographic~ screen}~{\rm (HS)}. \end{equation}

According to Ref.\cite{CCO}, the Raychaudhuri equation (\ref{nc2})
was obtained by considering both the equipartition rule (\ref{eq4e})
and the (Unruh) temperature \begin{equation}
T_*=\frac{a_r}{2\pi}=-\frac{r\ddot{R}}{2\pi}\end{equation}
 on the boundary screen $\partial {\cal V}$
enclosing the spatial region of volume ${\cal
V}=\frac{4}{3}\pi\hat{r}^3$ with $\hat{r}=rR(t)$. Here,
$a_r=-d^2\hat{r}/dt^2$ is considered as the physical acceleration
for a radial comoving observer located at a point of the boundary
screen. However, it is well-known that the proper acceleration
vanishes for a comoving observer in the relativistic FRW approach.
Hence it is strange to introduce the (Unruh) temperature $T_*$ in
the nonrelativistic approach.  Explicitly, in order to derive the
Raychaudhuri equation (\ref{nc2}), they have used the equipartition
relation directly
\begin{equation} \label{equt}
E(=M)=\frac{NT_*}{2}.
\end{equation}
However, the (Unruh) temperature $T_*$ was not proven to be valid
for this  case. Padbanabhan~\cite{Pad1} has shown that there is no
simple justifications for defining $T_*$ using the acceleration of
geodesic derivation vector and thus, his successes of obtaining the
Raychaudhuri equation must be considered as fortuitous.   This
suggests that the recovery of the Raychaudhuri equation from the
equipartition rule together with the temperature $T_*$ seems to be
accidental. Otherwise, one has to explain why the equipartition rule
with $T_*$ is equivalent to the Raychaudhuri equation for the dust
matter. This could not be explained in view of  the
(nonrelativistic) entropic force because the equipartition rule was
used to define the temperature on the holographic screen. We note
that the temperature $T_*$ is positive only for a deceleration of
matter-dominated universe. On the other hand, $T_*$ is negative for
an acceleration of $\omega <-1/3$, which shows that $T_*$  is not
acceptable when extending other matter contents. Although they were
succeeded in deriving the Friedmann equation (\ref{Friedmann}) from
the Raychaudhuri equation (\ref{Ray}) by replacing the source mass
$M$ by the Tolman-Komar mass and integrating the resulting equation,
it has nothing to do with a nonrelativistic force law of (\ref{eq2})
and  (\ref{eq3}) because they have used the equipartition rule
(\ref{equt}). Furthermore, it is not clear why the equipartition
rule does  provide  the Raychaudhuri equation which corresponds to
the Newtonian force law in the cosmological setting.

As a slightly different approach, the author~\cite{shek} has imposed
the Newtonian force law for a test particle $m$  near the boundary
screen $\partial {\cal V}$ directly
\begin{equation} \label{forcec}
F=m
r\ddot{R}=F_{ent},~~F_{ent}=-\frac{GMm}{R^2}\Bigg[1-\frac{\beta}{\pi}\frac{G}{R^2}-\frac{\gamma}{4\pi^2}\frac{G^2}{R^4}\Bigg]
\end{equation}
 to find the
corrected-Raychaudhuri equation by considering the corrected-entropy
\begin{equation}
S_c(A)=\frac{A}{4G}-\beta \ln \Big[\frac{A}{4G}\Big]+\gamma
\frac{G}{A}+\cdots. \end{equation} It seems that this approach
mimics  the Newtonian cosmology because (\ref{forcec}) is the same
as in (\ref{NFL}) when disregarding the correction terms. However,
the origin of the force ($F_{ent}$) is different from the Newtonian
cosmology because he has   used the equipartition rule and
holographic principle prior to the derivation of  the
corrected-entropic force $F_{ent}$ on the holographic screen.

The other is a derivation of Friedmann equation by using the
apparent horizon as the holographic screen and taking into account
the differential of equipartition rule (equivalently, the first law
of thermodynamics)~\cite{SG}. They have started with considering the
relativistic FRW metric to define the apparent horizon. Hence, this
is surely a relativistic approach.  As was previously emphasized, it
is unlikely to define the apparent horizon in the nonrelativistic
Newtonian cosmology.  This amounts to introducing the Schwarzschild
spacetime to define a proper holographic screen~\cite{MKent}.  If
one introduces the apparent horizon $r_A$ properly, thermodynamics
is defined by giving the Hawking temperature
\begin{equation}
T_A=\frac{1}{2\pi r_A},~r_A=\frac{1}{\sqrt{H^2+k/R^2}}
\end{equation}
with the Hubble parameter $H=\dot{R}/R$.  In this case, the
differential of the equipartition rule is equivalent to the first
law of thermodynamics as
\begin{equation} dE_A=\frac{N_A}{2}dT_A+\frac{T_A}{2}dN_A=\frac{dr_A}{G}
\longleftrightarrow dE_A=T_AdS_A \end{equation} where
\begin{equation}
N_A=\frac{4\pi r_A^2}{G}=4S_A.
\end{equation}
Hence, the equipartition rule is nothing new  but  redundant.
Supplying the perfect fluid as the energy-momentum  tensor
$T_{\mu\nu}=(\rho+p)u_\mu u_\nu + p g_{\mu\nu}$ like
\begin{equation}
dE_A=4\pi \tilde{r}_A^2 T_{\mu\nu} k^\mu k^\nu dt
\end{equation}
they have led to the Friedmann equation (\ref{Friedmann}) and then,
the Raychaudhuri equation (\ref{Ray}). However, this approach is not
a non-relativistic approach and thus has nothing to do with the
entropic force defined in (\ref{eq2}) and (\ref{eq3}), since the
Friedmann equation was derived from the first law of thermodynamics
(differential of equipartition rule but not equipartition rule
itself).

Finally, we do not mention the entropic force applications to the
accelerating universe~\cite{LWh,EFS} and the inflation~\cite{EFSi}
because these issues are beyond the Newtonian cosmology.

\section{Discussions}
First of all, we wish to point out the difference between Newtonian
gravity and general relativity~\cite{NHw}. Newtonian gravity is an
action-at-a-distance which means that  the gravitational influence
propagates instantaneously, implying the violation of causality.
Also, Newtonian gravity is ignorant of the presence of horizon where
the relativistic effects are supposed to dominates. For example, the
horizons are  considered as  either the event horizon of black hole
or the apparent horizons in the FRW universe. It is suggested  that
the absence of horizon does not enable to implement the holographic
screen in the Newtonian gravity (cosmology). The holographic
principle (holographic screen) appears in the black hole which is
formed after the gravitational collapse through the supernova
explosion and in the FRW universe based on the cosmological
principle when using the relativistic approach~\cite{SL}. In this
sense, the holographic principle seems to have nothing do to with
the Newtonian gravity (cosmology).

 It seems that the
entropic force (\ref{eq2}) based on the biophysics is not realized
in the Newtonian cosmology unless the holographic screen is
implemented. Actually, there is no justification (\ref{hsp}) of
taking the boundary surface ($\partial V$) enclosing an expanding
cavity $V$ as the holographic screen (HS) which contains space,
time, and information. Presumably, if  the holographic screen were
introduced in the Newtonian mechanics, one would propose the
equipartition rule on the holographic screen. This means that
gravitational attraction could be the result of the way that
information about material objects is organized in space.
Accordingly, we could define the entropic force to reproduce the
Newtonian force law. This is a way of realizing an entropic force as
an emergent phenomena which arises from the statistical behavior of
microscopic degrees of freedom encoded on  the holographic
screen~\cite{Ver}. However, this approach is too abstract to realize
the entropic force. Thus, the Verlinde's proposition on the entropic
force may be a hand-waving dimensional argument on the gravitational
(cosmological) side.

In conclusion, we have not confirmed   the connection between
Newtonian cosmology and  entropic force. We hope that the Newtonian
cosmology may provide a simple testbed to prove the entropic force.

\section*{Acknowledgment}
 This work was in part  supported  by Basic
Science Research Program through the National Research  Foundation
(NRF) of Korea funded by the Ministry of Education, Science and
Technology (NO.2010-0028080).

\end{document}